\documentclass[11pt]{article}

\usepackage{color}
\usepackage{amsmath}
\usepackage{graphicx}
\usepackage{amssymb}
\usepackage{amsfonts}
\usepackage{textcomp}

\newtheorem{theorem}{Theorem}
\newtheorem{lemma}{Lemma}

\title{Bifurcations of blowup in inviscid shell models \\ of convective turbulence}
\author{Alexei A. Mailybaev\thanks{Instituto 
Nacional de Matem\'atica Pura e Aplicada -- IMPA, Rio de Janeiro, Brazil. Email: alexei@impa.br}}

\begin{document}
\maketitle

\begin{abstract}
We analyze the blowup (finite-time singularity) in inviscid shell models of convective turbulence. We show that the blowup exists and its internal structure undergoes a series of bifurcations under a change of shell model parameter. Various blowup structures are observed and explained, which vary from self-similar to periodic, quasi-periodic and chaotic regimes. Though the blowup takes sophisticated forms, its asymptotic small-scale structure is independent of initial conditions, i.e., universal. Finally, we discuss implications of the obtained results for the open problems of blowup in inviscid flows and for the theory of turbulence.
\end{abstract}


\section{Introduction}
Blowup in a flow of incompressible ideal fluid, i.e., formation of a finite-time singularity in a regular solution of flow equations is an important phenomenon, which  may be responsible for the energy cascade in developed turbulence~\cite{frisch1995turbulence,eyink2006onsager,mailybaev2012computation}. The existence of blowup for fully inviscid 2D natural convection (Boussinesq equations) and for 3D incompressible Euler equations are the outstanding open problems of mathematical fluid dynamics~\cite{chae2008incompressible,gibbon2008three,gibbon2008three2}. Dynamical (shell) models provide essential simplification of the flow equations, where the Fourier space is substituted by a sequence of discrete wave numbers increasing in geometric progression. 
Despite of simplicity, such models preserve many nontrivial features of the original system like, e.g., energy and entropy cascades and anomalous scaling of turbulent spectra~\cite{biferale2003shell}. Shell models allow very precise numerical simulation. This motivates the study of these models with the hope to get a clue for explanation of turbulent phenomena. However, turbulent dynamics in shell models is still not well understood and explained. 

Mathematical formalism for the Gledzer-Ohkitani-Yamada (GOY)~\cite{gledzer1973system,ohkitani1989temporal} and Sabra~\cite{l1998improved} inviscid shell models was developed in \cite{barbato2006some,constantin2007regularity}, where global existence of weak solutions was proved and the criterion for blowup of strong solutions was given. 
For the dyadic shell model the finite time blowup  
can be proved rigorously~\cite{katz2005finite,cheskidov2007inviscid}.
A constructive method for analysis of blowup structure in inviscid shell models was proposed in~\cite{dombre1998intermittency}. This analysis uses renormalized variables, where the blowup time is shifted to infinity. Traveling wave solutions of the renormalized system lead to the blowup with self-similar structure. Such self-similar blowup's were observed numerically in different shell and cascade models of inviscid flows~\cite{siggia1978model,nakano1988,uhlig1997singularities,l2001outliers}.
A similar renormalization technique can be introduced for continuous systems~\cite{mailybaev2012}, see also~\cite{eggers2009role} discussing self-similarity of singularities in partial differential equations. 

In this paper, we consider the shell model of natural convection proposed in~\cite{brandenburg1992energy}. Properties of this model in the turbulent regime were studied, e.g., in~\cite{suzuki1995entropy,ching2002statistically,ching2008anomalous,Ching2010}.  
In particular, it reproduces the Bolgiano-Obukhov 
scaling law~\cite{siggia1994high,lohse2010small} of turbulent spectra in the inertial range (with anomalous corrections). 
Our goal is to study the blowup in a family of fully inviscid models with a parameter controlling the energy transfer between shells.

First we derive the blowup criterion for the inviscid shell model, 
which is similar to the Beale-Kato-Majda theorem for incompressible Euler 
equations~\cite{beale1984remarks}. Then the renormalization scheme is given generalizing 
the one used in~\cite{dombre1998intermittency,mailybaev2012}. 
Our new contribution is the introduction of the Poincar\'e map for the renormalized system. 
This approach allows rigorous analysis relating blowup structures to attractors of 
the Poincar\'e map. In particular, the self-similar blowup structures detected earlier 
correspond to fixed-point attractors. 
With a change of parameter of the shell model a series of bifurcations is observed, 
which includes a period doubling cascade to chaos and provides a large variety of attractors. 
We show that all types of attractors (fixed-point, periodic, quasi-periodic and chaotic) 
describe universal asymptotic blowup structures. For each type of attractor, 
we give the detailed description of blowup and determine its scaling properties. 

The obtained results give some insight to the open problems of blowup in 
continuous inviscid flow models. We see that the blowup can 
have the universal asymptotic structure, which does not depend on initial 
conditions up to system symmetries. However, this structure does not 
have to be self-similar and a number of different scenarios including periodic 
and chaotic regimes are possible. Thus, it may be very hard to detect such universality, 
unless special renormalization tools are applied like the one proposed in this work.

The paper is organized as follows. In Section~2 we introduce the model and give 
the blowup criterion. Section~3 describes the renormalization scheme of~\cite{dombre1998intermittency}, 
which is demonstrated on the isothermal shell model in Section~4. 
The Poincar\'e map for renormalized dynamics is defined in Section~5 and the corresponding bifurcation diagram is studied in Section~6. Sections~7--9 analyze blowup structures corresponding to fixed-point, periodic, quasi-periodic and chaotic attractors of the Poincar\'e map. In Section 10 we summarize the results and discuss their relation to the open problems of blowup in continuous models and anomalous spectra of convective turbulence.

\section{Blowup in inviscid models of natural convection}

In the Boussinesq approximation~\cite{landau1987course}, buoyancy-driven flows 
of unit density are governed by the equations
\begin{equation}
\frac{\partial\mathbf{u}}{\partial t}
= -\mathbf{u}\cdot\nabla\mathbf{u}
-\nabla p+\nu\Delta\mathbf{u}+\alpha g\theta\mathbf{e}_z,
\quad
\frac{\partial\theta}{\partial t}
= -\mathbf{u}\cdot\nabla\theta
+\kappa\Delta\theta,
\quad \nabla\cdot\mathbf{u} = 0.
\label{eq2.1} 
\end{equation}
Here $\mathbf{u}$ is the velocity, $p$ is the pressure, and $\theta = T-T_0$ is the deviation of temperature $T$ from the mean value $T_0$. The constant parameters $\kappa$, $\nu$, $\alpha$, and $g$ are the thermal diffusivity, the kinetic viscosity, the thermal expansion coefficient, and the gravitational acceleration, respectively; the unit vector $\mathbf{e}_z$ specifies the vertical direction. In the inviscid limit, the system conserves the entropy $S = \int \frac{1}{2}\theta^2dV$ and the total energy $E = \int (\frac{1}{2}|\mathbf{u}|^2+\alpha g z\theta)dV$, where $z$ is the vertical coordinate. 

Our work is motivated by the problem of blowup of classical solutions for the fully inviscid system, i.e., formation of a finite-time singularity in smooth solutions when $\nu = \kappa = 0$. 
In the isothermal case, $\theta = 0$, inviscid equations (\ref{eq2.1}) reduce to the incompressible Euler equations. 
It is well-known that the Euler equations for ideal incompressible fluid in 2D possess a global in time, unique,
regular solution, see, e.g., \cite{majda2001vorticity}. Global regularity of the 3D incompressible Euler equations is an open problem~\cite{chae2008incompressible,gibbon2008three,gibbon2008three2}. The  Beale-Kato-Majda theorem~\cite{beale1984remarks} states that, if the initially smooth solution of the Euler equations cannot be continued beyond the time $t_c$ and $t_c$ is the first such time, then
\begin{equation}
\int_0^{t_c} \|\boldsymbol\omega(\cdot,t)\|_\infty dt 
= \infty,
\label{eq2.1b} 
\end{equation}
where $\|\boldsymbol\omega(\cdot,t)\|_\infty$ denotes the $L^\infty$ norm 
of the vorticity $\boldsymbol\omega = \nabla\times\mathbf{v}$ at time $t$. 
This condition implies that the maximum vorticity must grow at least as $(t_c-t)^{-1}$ near the blowup.
In the general case $\theta \ne 0$, the existence of blowup in smooth solutions 
of the fully inviscid Boussinesq system (\ref{eq2.1}) is an open problem both for 2D and 3D 
flows~\cite{chae2008incompressible}. The blowup criterion for 2D flows 
includes (\ref{eq2.1b}) and an extra condition for the 
temperature gradient~\cite{weinan1994small,chae1997local}.

Dynamical models, also called shell models, represent an essential simplification 
of system (\ref{eq2.1}) constructed in the discretized Fourier space $k_n = k_0h^n$ with $h > 1$. 
Here $k_n$ with $n = 0,1,2,\ldots$ is the wavenumber, and the corresponding effective velocity 
and temperature are given by real quantities $u_n$ and $\theta_n$. 
We will consider the shell model proposed in \cite{brandenburg1992energy}, which has the form
\begin{eqnarray}
\frac{du_n}{dt} & = & Ak_n(u_{n-1}^2-hu_nu_{n+1})+Bk_n(u_nu_{n-1}-hu_{n+1}^2)
-\nu k_n^2 u_n+\alpha g\theta_n+f_n,
\label{eq2.2}
\\[3pt]
\frac{d\theta_n}{dt} & = & \widetilde{A}k_n(u_{n-1}\theta_{n-1}-hu_n\theta_{n+1})
+\widetilde{B}k_n(u_n\theta_{n-1}-hu_{n+1}\theta_{n+1})
-\kappa k_n^2 \theta_n+g_n,
\label{eq2.3}
\end{eqnarray}
where $A$, $B$, $\widetilde{A}$, and $\widetilde{B}$ are real parameters and 
$u_{-1} = \theta_{-1} = 0$. 
Equations (\ref{eq2.2}) and (\ref{eq2.3}) have the structure similar to the 
Fourier transformed Boussinesq equations (\ref{eq2.1}). Nonlinear quadratic terms 
describe the shells interaction and involve the $n$th 
and its neighboring shells. 
The additional forcing terms $f_n$ and $g_n$ model the influence of boundaries 
(e.g., due to heating from below) and are usually restricted to the first shells. 
Despite of simplicity, the shell models demonstrate several highly nontrivial 
properties of the original continuous system (\ref{eq2.1}), for example, the turbulent 
regime with entropy cascade and the anomalous statistics close to the Bolgiano-Obukhov 
scaling law in the inertial range~\cite{brandenburg1992energy,suzuki1995entropy}.

We assume the traditional choice of parameters, 
$h = 2$, $k_0 = \alpha g = B = \widetilde{A} = \widetilde{B} = 1$, and consider $A = \varepsilon$. The parameter $\varepsilon$ controls the nonlinear process of energy transfer between shells, and the value $\varepsilon = 0.01$ was chosen in earlier works, e.g., \cite{brandenburg1992energy,Ching2010}. We will study the fully inviscid equations (\ref{eq2.2}), (\ref{eq2.3}) with no forcing, which reduce to
\begin{eqnarray}
\frac{du_n}{dt} & = & k_n[\varepsilon(u_{n-1}^2-hu_nu_{n+1})+u_nu_{n-1}-hu_{n+1}^2]
+\theta_n,
\label{eq2.4}
\\[3pt]
\frac{d\theta_n}{dt} & = & 
k_n\left(u_{n-1}\theta_{n-1}-hu_n\theta_{n+1}
+u_n\theta_{n-1}-hu_{n+1}\theta_{n+1}\right).
\label{eq2.5}
\end{eqnarray}
This system conserves the entropy $S = \sum \frac{1}{2}\theta_n^2$ 
(here and below the sums are taken over all shells $n$). 
If $\theta_n \equiv 0$, the energy $E = \sum \frac{1}{2}u_n^2$ is conserved. 
However, there is no analog for the total energy in the case $\theta_n \ne 0$ for this shell model.

Following~\cite{constantin2007regularity}, we introduce the norms
\begin{equation}
\|u\|_1 := \left(\sum k_n^{2}u_n^2\right)^{1/2},\quad 
\|u\|_{1,\infty} := \sup_n k_n|u_n|, 
\label{eq51}
\end{equation}
and consider smooth solutions with finite norms
\begin{equation}
\|u\|_1+\|\theta\|_1 < \infty. 
\label{eq51b}
\end{equation}
Local existence of such solutions can be proved using the Picard theorem 
in the same way as for the inviscid Sabra shell model 
\cite{constantin2007regularity} with obvious modifications due to different nonlinear terms.
The blowup at $t = t_c$ implies that 
\begin{equation}
\sup_{0\le t < t_c}\left(\|u\|_1+\|\theta\|_1\right) 
= \infty.
\label{eq51c}
\end{equation}
Note that the blowup does not imply a singularity for a 
particular shell variable $u_n$ or $\theta_n$, since they represent Fourier components of the flow; only the integral contribution (the norm) explodes. 
The following theorem provides the blowup criterion, which is similar to (\ref{eq2.1b}) and to the one derived in \cite{constantin2007regularity} for 
the Sabra shell model.

\begin{theorem}
Let $u_n(t)$ and $\theta_n(t)$, $0 \le t < t_c$ be a smooth solution of the inviscid shell model 
(\ref{eq2.4}), (\ref{eq2.5}) satisfying condition (\ref{eq51b}), 
where $t_c$ is its maximal time of existence. Then, either $t_c = \infty$ or 
\begin{equation}
\int_0^{t_c} \|u\|_{1,\infty} dt  = \infty.
\label{eq52}
\end{equation}
\label{th1}
\end{theorem}

\noindent\textit{Proof:} 
Condition (\ref{eq52}) implies that $\|u\|_{1,\infty}$ is unbounded for $0 \le t < t_c$ and, hence, condition (\ref{eq51c}) is satisfied. Therefore, (\ref{eq52}) is a sufficient condition for the blowup.
 
Using (\ref{eq2.4}), (\ref{eq2.5}) and the norm defined in (\ref{eq51}), we derive
\begin{equation}
\begin{array}{rcl}
\displaystyle
\frac{1}{2}\frac{d}{dt}\left(\|u\|_1^2+\|\theta\|_1^2\right) 
& = & 
\displaystyle
\sum u_nk_n^{3}\left[\varepsilon(u_{n-1}^2-hu_nu_{n+1})
+u_nu_{n-1}-hu_{n+1}^2\right]
+\sum k_n^{2}u_n\theta_n
\\[10pt]
& & 
\displaystyle
+\sum\theta_nk_n^{3}\left(u_{n-1}\theta_{n-1}-hu_n\theta_{n+1}
+u_n\theta_{n-1}-hu_{n+1}\theta_{n+1}\right).
\end{array}
\label{eq53}
\end{equation}
Using the condition $k_n|u_n| \le \|u\|_{1,\infty}$ and 
applying Cauchy-Schwarz inequality in (\ref{eq53}) yields
\begin{equation}
\frac{d}{dt}\left(\|u\|_1^2+\|\theta\|_1^2\right) 
\le C\|u\|_{1,\infty}\|u\|_1^2+2\|u\|_1\|\theta\|_1
+C\|u\|_{1,\infty}\|\theta\|_1^2
\le \left(1+C\|u\|_{1,\infty}\right)\left(\|u\|_1^2+\|\theta\|_1^2\right)
\label{eq54}
\end{equation}
valid for a sufficiently large positive constant $C$. Using the Gronwall's inequality we get
\begin{equation}
\left(\|u\|_1^2+\|\theta\|_1^2\right)_{t = t_c} 
\le \left(\|u\|_1^2+\|\theta\|_1^2\right)_{t = 0}
\exp\left(t_c+C\int_0^{t_c}
\|u\|_{1,\infty}dt\right).
\label{eq55}
\end{equation}
This proves that condition (\ref{eq52}) is necessary for the blowup. $\square$

Condition (\ref{eq52}) implies that the maximum shell ``vorticity'' must grow at least as $k_nu_n \sim (t_c-t)^{-1}$ as $t \to t_c^-$. It is interesting that no extra condition for the shell temperatures is necessary. We remark that the statements of this section (on the local existence and blowup) remain valid for the sequences with $n \in \mathbb{Z}$, i.e., without the left end at $n = 0$. Also, the norm $\|u\|_1$ can be replaced by $\|u\|_d = \left(\sum k_n^{2d}u_n^2\right)^{1/2}$ with $d > 1$ for the sequences with $n \ge 0$, but not for $n \in \mathbb{Z}$.

\section{Dombre-Gilson renormalization scheme}

In this section we introduce the renormalization scheme analogous to the one suggested by Dombre and Gilson~\cite{dombre1998intermittency} for the mixed Obukhov-Novikov (ON) model~\cite{obukhov1971some,desnyansky1974evolution}. 
The aim of this renormalization is to obtain a system, where the blowup time $t_c$ is moved to infinity, so that standard dynamical system methods can be applied.  

The new time variable $\tau$ is defined implicitly as 
\begin{equation}
t = \int_0^\tau \exp\left(-\int_0^{\tau'} A(\tau'') d\tau''
\right)d\tau',
\label{eq9}
\end{equation}
where the function $A(\tau)$ will be specified later. The scaled speed $v_n$ and temperature $\vartheta_n$ are given by
\begin{equation}
v_n = \exp\left(-\int_0^\tau A(\tau') d\tau'\right)k_nu_n,\quad
\vartheta_n = \exp\left(-2\int_0^\tau A(\tau') d\tau'\right)k_n\theta_n.
\label{eq10}
\end{equation}
Differentiating expressions (\ref{eq10}) with respect to $\tau$ and using (\ref{eq2.4}), (\ref{eq2.5}) and (\ref{eq9}) yields the system
\begin{eqnarray}
\frac{dv_n}{d\tau} & = & P_n-Av_n,\quad 
P_n = \varepsilon(h^2v_{n-1}^2-v_nv_{n+1})
+hv_nv_{n-1}-h^{-1}v_{n+1}^2+\vartheta_n,
\label{eq7}
\\[3pt]
\frac{d\vartheta_n}{d\tau} & = & 
Q_n-2A\vartheta_n,\quad
Q_n = h^2v_{n-1}\vartheta_{n-1}-v_n\vartheta_{n+1}
+hv_n\vartheta_{n-1}-h^{-1}v_{n+1}\vartheta_{n+1}.
\label{eq8}
\end{eqnarray}
It is easy to see that 
\begin{equation}
\frac{d}{d\tau}\sum (v_n^4+\vartheta_n^2) = 
\sum (4v_n^3P_n+2\vartheta_nQ_n)-4A\sum (v_n^4+\vartheta_n^2).
\label{eq11}
\end{equation}
Let us choose
\begin{equation}
A(\tau) = \sum \left(v_n^3P_n+\frac{1}{2}\vartheta_nQ_n\right)\left/
\sum \left(v_n^4+\vartheta_n^2\right)\right..
\label{eq12}
\end{equation}
Then the sum 
\begin{equation}
\sum (v_n^4+\vartheta_n^2) = c 
\label{eq12b}
\end{equation}
is conserved. At $\tau = 0$ expressions (\ref{eq10}) and (\ref{eq51}), (\ref{eq51b}) yield $\sum v_n^4 \le \left(\sum v_n^2\right)^2 = \|u\|_1^4 < \infty$ and $\sum \vartheta_n^2 = \|\theta\|_1^2 < \infty$. Hence, the sum in (\ref{eq12b}) is finite, $c < \infty$. In the next lemma we show that the function in (\ref{eq12}) is well defined for any nontrivial solution (we will not consider the trivial solution $v_n \equiv 0$, $\vartheta_n \equiv 0$ from now on).

\begin{lemma}
Nontrivial solution $v_n(\tau)$, $\vartheta_n(\tau)$ of system (\ref{eq7}), (\ref{eq8}) exists globally for $0 \le \tau < \infty$ and it is related by (\ref{eq9}), (\ref{eq10}) to  the solution $u_n(t)$, $\theta_n(t)$ of system (\ref{eq2.4}), (\ref{eq2.5}) with  $t < t_c$, where $t_c$ is the blowup time from Theorem~\ref{th1}. 
\label{lemma1}
\end{lemma}

\noindent\textit{Proof:}
Because of relations (\ref{eq9}), (\ref{eq10}) between the solutions, we need to show only that $A(\tau)$ in (\ref{eq12}) is well defined and that any finite $\tau \ge 0$ corresponds to $t < t_c$. 
Due to (\ref{eq12b}), we have
\begin{equation}
|v_n| < c^{1/4}.
\label{eq12c}
\end{equation}
The denominator in (\ref{eq12}) is constant. We substitute $P_n$, $Q_n$ from (\ref{eq7}), (\ref{eq8}) into (\ref{eq12}) and consider the terms in the nominator. The sum of first terms can be bounded as
\begin{equation}
\left|\varepsilon h^2 \sum v_n^3v_{n-1}^2\right| \le 
|\varepsilon|h^2c^{1/4}\sum v_n^2v_{n-1}^2 \le
|\varepsilon|h^2c^{1/4}\sum v_n^4 \le
|\varepsilon|h^2c^{5/4},
\label{eq12d}
\end{equation}
where we used (\ref{eq12c}), the Cauchy-Schwarz inequality and (\ref{eq12b}).
Similar bounds can be found for the other terms. It follows that $|A(\tau)|$ is bounded for all $\tau \ge 0$. 

Using (\ref{eq12c}) and finiteness of the integral in (\ref{eq10}), one can show that the sequence $|k_nu_n(t)|$ is bounded for any fixed $t$ corresponding to $0 \le \tau < \infty$. Hence, $\|u\|_{1,\infty} < \infty$ and, by Theorem~\ref{th1}, we have $t < t_c$.
$\square$

Equations (\ref{eq7}), (\ref{eq8}) with $A$ from (\ref{eq12}) possess three types of symmetries  
\begin{eqnarray}
(i)&& 
\tau \mapsto \tau/a,\quad v_n \mapsto av_n,\quad \vartheta_n \mapsto a^2\vartheta_n; 
\label{eq15}
\\[5pt]
(ii)&&  
\tau \mapsto \tau-\tau_0;
\label{eqSym}
\\[5pt]
(iii) &&  
v_n \mapsto v_{n+1},\quad \vartheta_n \mapsto \vartheta_{n+1}.
\label{eqSym2}
\end{eqnarray}
The last symmetry does not hold for the equations of the boundary shell $n = 0$, 
but this symmetry becomes exact when we pass to the infinite lattice $n \in \mathbb{Z}$ in Section~\ref{secPoinc}. 

\begin{lemma}
The symmetries (\ref{eq15})--(\ref{eqSym2}) are equivalent to the following symmetries for the original shell model (\ref{eq2.4}), (\ref{eq2.5}): 
\begin{eqnarray}
(i)&& 
t \mapsto t/a,\quad u_n \mapsto au_n,\quad \theta_n \mapsto a^2\theta_n; 
\label{eq15n}
\\[5pt]
(ii)&&  
t \mapsto t/a-t_0, \quad u_n \mapsto au_n,\quad \theta_n \mapsto a^2\theta_n;
\label{eqSymn}
\\[5pt]
(iii) &&  
u_n \mapsto hu_{n+1},\quad \theta_n \mapsto h\theta_{n+1}.
\label{eqSym2n}
\end{eqnarray}
In the case $(ii)$ the values of $t_0$ and $a$ are uniquely determined by $\tau_0$.
\label{lemma2}
\end{lemma}

\noindent\textit{Proof:}
Symmetries (\ref{eq15n})--(\ref{eqSym2n}) describe time scaling, time shift and space scaling (recall that $k_{n+1} = hk_n$), and can be related to (\ref{eq15})--(\ref{eqSym2}) using (\ref{eq9}) and (\ref{eq10}). We will consider in detail the most difficult case $(ii)$. The symmetries $(i)$ and $(iii)$ are studied similarly.

According to (\ref{eqSym}), the new variables (denoted by a hat) are 
\begin{equation}
\widehat\tau = \tau-\tau_0,\quad 
\widehat{v}_n(\widehat{\tau}) = v_n(\tau),\quad 
\widehat{\vartheta}_n(\widehat{\tau}) = \vartheta_n(\tau).
\label{eq80}
\end{equation}
It follows from (\ref{eq12}) that 
\begin{equation}
\widehat{A}(\widehat\tau) = A(\tau) = A(\widehat\tau+\tau_0). 
\label{eq81}
\end{equation}
For the time $\widehat{t}$ corresponding to $\widehat{\tau}$, expression (\ref{eq9}) yields
\begin{equation}
\begin{array}{rcl}
\widehat{t} 
& = & \displaystyle 
\int_0^{\widehat{\tau}} 
\exp\left(-\int_0^{\tau'} \widehat{A}(\tau'') d\tau''\right)d\tau'
= \int_0^{\tau-\tau_0} 
\exp\left(-\int_0^{\tau'} A(\tau''+\tau_0) d\tau''\right)d\tau'
\\[15pt]
& = & \displaystyle 
\int_{\tau_0}^{\tau} 
\exp\left(-\int_{\tau_0}^{\xi'} A(\xi'') d\xi''\right)d\xi',
\end{array}
\label{eq82}
\end{equation}
where we used $\widehat\tau = \tau-\tau_0$ from (\ref{eq80}), $\widehat{A}$ from (\ref{eq81}) and made the changes of variables $\xi' = \tau'+\tau_0$ and $\xi'' = \tau''+\tau_0$.
Expression (\ref{eq82}) can be written as
\begin{equation}
\widehat{t} = t/a-t_0,
\label{eq83}
\end{equation}
where
\begin{equation}
a = \exp\left(-\int_0^{\tau_0} A(\tau'') d\tau''\right),\quad
t_0 = \int_0^{\tau_0}\exp\left(-\int_0^{\tau'} A(\tau'') d\tau''\right)d\tau'
\label{eq83b}
\end{equation}
and $t$ is given by (\ref{eq9}). For the shell speeds, the first expression in (\ref{eq10}) with (\ref{eq80}) and (\ref{eq81}) yield
\begin{equation}
\widehat{u}_n(\widehat{t}) 
= \exp\left(\int_0^{\widehat\tau} \widehat{A}(\tau') d\tau'\right)k_n^{-1}\widehat{v}_n(\widehat\tau)
= \exp\left(\int_0^{\tau-\tau_0} A(\tau'+\tau_0) d\tau'\right)k_n^{-1}
v_n(\tau).
\label{eq84}
\end{equation}
The change of variable $\xi' = \tau'+\tau_0$ in the integral yields
\begin{equation}
\widehat{u}_n(\widehat{t}) 
= \exp\left(\int_{\tau_0}^{\tau} A(\tau') d\tau'\right)k_n^{-1}
v_n(\tau) = au_n(t)
\label{eq85}
\end{equation}
with $a$ from (\ref{eq83b}).
Similarly, one derives $\widehat{\theta}_n(\widehat{t}) = a^2\theta_n(t)$ using the second expression in (\ref{eq10}), which accomplishes the proof of (\ref{eqSymn}). $\square$

\section{Universal structure of blowup}
\label{sec5}

\begin{figure}
\centering \includegraphics[width = 0.98\textwidth]{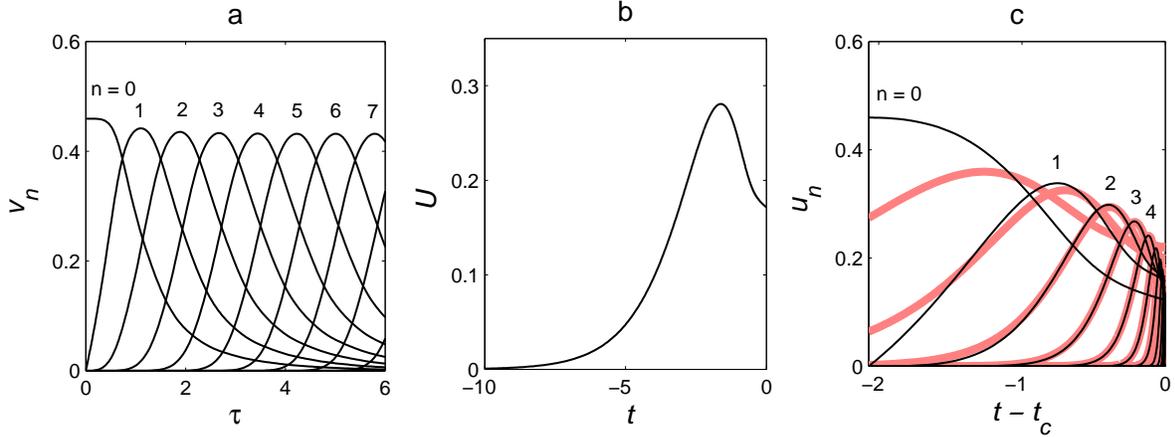}
\caption{Blowup in isothermal (mixed ON) shell model: (a) traveling wave in renormalized variables $v_n(\tau)$, (b) universal function $U(t)$, (c) numerical simulation of shell velocities $u_n(t)$ (black curves) and their asymptotic self-similar form (bold light-red curves). The blowup time $t_c = 2.07$ corresponds to $\tau \to \infty$.}
\label{fig1}
\end{figure}

As a demonstration of the renormalization method, let us consider the isothermal (mixed ON) shell model with $\theta_n = \vartheta_n = 0$ and $\varepsilon = 0.5$. 
Numerical solution of equations (\ref{eq7})  
is shown in Fig.~\ref{fig1}a. 
Here we used 100 shells and the initial condition with only the first shell speed perturbed, $v_0(0) > 0$.
For large times, the asymptotic solution has the form
\begin{equation}
v_n(\tau) = aV(n-a\tau),
\label{eq13}
\end{equation}
which was first observed in~\cite{dombre1998intermittency}. 
Expression (\ref{eq13}) defines a solitary wave traveling to larger shell numbers $n$ with constant speed $a$. The function $V(\xi)$ vanishes in the limits $\xi \to \pm\infty$, and the constant $a > 0$ is arbitrary reflecting the scaling symmetry (\ref{eq15}). 

The theory explaining asymptotic solution (\ref{eq13}) will be given 
in the next sections using the dynamical system approach. This solution yields the self-similar expression for original variables $u_n(t)$. 
In the next theorem we provide such self-similar expression and its derivation (in modified form) given by Dombre and Gilson; we will need these results later for the study of nonzero shell temperatures.
Note that analogous self-similar blowup was observed numerically in 
\cite{siggia1978model,nakano1988,uhlig1997singularities,l2001outliers}
for different dynamical models of hydrodynamic flows. 

\begin{theorem}[\cite{dombre1998intermittency}]
Taking $a = 1$ in (\ref{eq13}) we define the real quantity
\begin{equation}
y = \frac{1}{\log h}\int_0^{1}A(\tau) d\tau
\label{eq17}
\end{equation}
and the function
\begin{equation}
U(t-t_c) = \exp\left(\int_0^\tau A(\tau')d\tau'\right)V(-\tau),
\label{eq22}
\end{equation}
where $\tau$ is related to $t$ implicitly by (\ref{eq9}) and $A$ is given by (\ref{eq12}). If $y > 0$, then the solution $u_n(t)$ corresponding to (\ref{eq13}) has the form (for arbitrary $a > 0$) 
\begin{equation}
u_n(t) = ak_n^{y-1}U(ak_n^y(t-t_c)), 
\label{eq21}
\end{equation}
where 
\begin{equation}
t_c = \int_0^\infty \exp\left(-\int_0^{\tau'} A(\tau'') d\tau''
\right)d\tau' < \infty
\label{eq19}
\end{equation}
is the blowup time.
\label{th2}
\end{theorem}

\noindent\textit{Proof:}
Let us consider the time-scaling symmetry (\ref{eq15}). By Lemma~\ref{lemma2}, the corresponding quantities $t$, $t_c$ and $u_n$ scale as
\begin{equation}
t-t_c \mapsto (t-t_c)/a,\quad u_n \mapsto au_n.
\label{eq15b}
\end{equation}
This shows that dependence on $a$ in (\ref{eq13}) and (\ref{eq21}) is the result of the same symmetry. Hence, we can take $a = 1$ in the rest of the proof. 

The function $A(\tau)$ given by (\ref{eq12}) and (\ref{eq13}) is periodic  with period $1/a = 1$. Then, it follows from (\ref{eq17}) that the inequality 
\begin{equation}
\int_0^{\tau}A(\tau) d\tau > C+\tau y\log h
\label{eq18}
\end{equation}
holds for some constant $C$. Using this estimate, 
it is straightforward to show that  the integral in (\ref{eq19}) converges for $y > 0$.

Due to periodicity of $A(\tau)$ and (\ref{eq17}) we have
\begin{equation}
\exp\left(\int_\tau^{\tau+n}A(\tau') d\tau'\right) = k_n^y
\label{eq17c}
\end{equation}
for any $\tau$, where $k_n = h^n$.
Let us consider the time $t'$ corresponding to $\tau+n$. 
Using (\ref{eq9}) and (\ref{eq19}), we obtain
\begin{equation}
\begin{array}{rcl}
t_c-t' 
& = & \displaystyle
\int_{\tau+n}^\infty \exp\left(-\int_0^{\tau'} A(\tau'') d\tau''
\right)d\tau'
=
\int_{\tau}^\infty \exp\left(-\int_0^{\widetilde{\tau}+n} A(\tau'') d\tau''
\right)d\widetilde{\tau}
\\[15pt]
& = & \displaystyle
k_n^{-y}\int_{\tau}^\infty \exp\left(-\int_0^{\widetilde{\tau}} A(\tau'') d\tau''
\right)d\widetilde{\tau}
= k_n^{-y}(t_c-t),
\end{array}
\label{eq34}
\end{equation}
where we changed the integration variable $\tau' = \widetilde{\tau}+n$ and then used (\ref{eq17c}). 
Similarly, we express the shell speed $u_n$ from (\ref{eq10}) and use (\ref{eq17c}), (\ref{eq13}) and (\ref{eq22}) as
\begin{equation}
u_n(t') 
= k_n^{-1}\exp\left(\int_0^{\tau+n} A(\tau') d\tau'\right)v_n(\tau+n) 
= k_n^{y-1}\exp\left(\int_0^{\tau} A(\tau') d\tau'\right)V(-\tau) = k_n^{y-1}U(t-t_c).
\label{eq33}
\end{equation}
Substituting $t-t_c = k_n^y(t'-t_c)$ expressed from (\ref{eq34}) into (\ref{eq33}) and dropping the prime, we obtain relation (\ref{eq21}) with $a = 1$.

Finally, let us show that expression (\ref{eq21}) describes the blowup at $t_c$. 
For this purpose we consider the times $t_n = t_c-bk_n^{-y}$ with arbitrary $b > 0$ such that $U(-b) \ne 0$. Then expression (\ref{eq21}) with $a = 1$ yields $k_nu_n(t_n) = k_n^yU(-b)$. For $y > 0$ we have $k_n|u_n(t_n)| \to \infty$ as $n \to \infty$ and, hence, the norm $\|u\|_1 \to \infty$. Since $t_n \to t_c$, this means the blowup at $t_c$. 
$\square$

Since (\ref{eq13}) is an asymptotic solution for the renormalized system, expression (\ref{eq21}) represents the asymptotic form of blowup for large $n$ and $t \to t_c^-$.
In our numerical example, computations yield $y = 0.8557$ and the function $U(t)$ is shown in Fig.~\ref{fig1}b. 
Figure~\ref{fig1}c compares the asymptotic self-similar relation (\ref{eq21}) 
with the numerical shell speeds $u_n(t)$ demonstrating convergence for large $n$. 
The scaling exponent $y$ and the function $U(t)$ in expression (\ref{eq21}) do not 
depend on initial conditions, which was confirmed numerically and will be explained theoretically in Section~\ref{sec_slefsim}. Therefore, the blowup has the universal self-similar asymptotic form. 

\section{Poincar\'e map}
\label{secPoinc}

Asymptotic traveling wave solutions (\ref{eq13}) do not exist for all values 
of parameter $\varepsilon$. Instead, solutions of system (\ref{eq7}), (\ref{eq8}) 
appear in the form of periodically or chaotically pulsating waves. 
We propose a general method for analysis of such solutions 
following the dynamical system approach and using 
the Poincar\'e map introduced below. 

Let us consider the infinite-dimensional space $W$ of renormalized flow variables 
\begin{equation}
\mathbf{w} = (\ldots,v_{n-1},v_n,v_{n+1},\ldots,
\vartheta_{n-1},\vartheta_n,\vartheta_{n+1},\ldots) \in W
\label{eq24}
\end{equation}
with the $\ell^2$ norm $\|\mathbf{w}\|^2 = \sum(v_n^2+\vartheta_n^2)$. 
According to (\ref{eq10}) and (\ref{eq51}), we have $\|\mathbf{w}\|^2 = c_u\|u\|_1^2+c_\theta\|\theta\|_1^2$, where the norms and the coefficients $c_u$, $c_\theta$ are finite for any $\tau \ge 0$ by Lemma~\ref{lemma1}. Hence, $\mathbf{w}(\tau) \in W$  for $\tau \ge 0$. 

The blowup phenomenon is described by dynamics for large shell numbers $n$. 
Hence, we can disregard the left end ($n = 0$) of the shell sequence in the blowup analysis, 
and consider equations (\ref{eq7}), (\ref{eq8}) for the sequences with $n \in \mathbb{Z}$. It is easy to check that all the statements made above about the shell model solutions remain valid in the case $n \in \mathbb{Z}$.

We define the real number 
\begin{equation}
n_v(\tau) = \sum n\left(v_n^4+\vartheta_n^2\right)\left/
\sum \left(v_n^4+\vartheta_n^2\right)\right.,
\label{eq23}
\end{equation}
which estimates the shell reached by the flow at time $\tau$. Recall that the denominator in (\ref{eq23}) is constant, see (\ref{eq12b}). 
Let us consider $\mathbf{w} = \mathbf{w}(0)$ as the initial condition for $\tau = 0$. Then we define the vector $\mathbf{w}' = \mathbf{w}(\tau_1)$ as the solution at the first time moment $\tau_1 > 0$ when the position $n_v$ increases by one, i.e.,
\begin{equation}
n_v(\tau_1) = n_v(0)+1.
\label{eq26}
\end{equation}
The corresponding transfer operator is denoted by $\mathcal{T}_1$, so that
\begin{equation}
\mathcal{T}_1\mathbf{w} = \mathbf{w}',\quad
v'_n = v_n(\tau_1),\quad \vartheta'_n = \vartheta_n(\tau_1). 
\label{eq28}
\end{equation}
The operator $\mathcal{T}_1$ is defined for given $\mathbf{w}$ if condition (\ref{eq26}) is satisfied for some $\tau_1 > 0$.
Numerical computations presented below suggest that this operator is defined for any nontrivial vector $\mathbf{w} \in W$, i.e., for any nontrivial solution we have $n_v \to \infty$ as $\tau \to \infty$. 

Also, we define the shift operator $\mathcal{S}: W \mapsto W$ acting as
\begin{equation}
\mathcal{S}\mathbf{w} = \mathbf{w}',\quad 
v'_n = v_{n+1},\quad \vartheta'_n = \vartheta_{n+1}.
\label{eq29}
\end{equation}
This operator represents the renormalization in physical space, because $n+1$ corresponds to the wave numbers $k_{n+1} = hk_n$.   

Now we can define the Poincar\'e map as $\mathcal{P} = \mathcal{S}\mathcal{T}_1$, i.e.,
\begin{equation}
\mathcal{P}\mathbf{w} = \mathbf{w}', \quad 
v'_n = v_{n+1}(\tau_1),\quad 
\vartheta'_n = \vartheta_{n+1}(\tau_1).
\label{eq30b}
\end{equation}
Therefore, the Poincar\'e map $\mathcal{P}$ represents the time-transfer 
operator combined with the shift of shell numbers. Due to this shift,  
the Poincar\'e map describes the flow dynamics observed 
in the ``moving frame'' along the logarithmic axis $n = \log_h k_n$ in the Fourier space.

\begin{figure}
\centering \includegraphics[width = 0.8\textwidth]{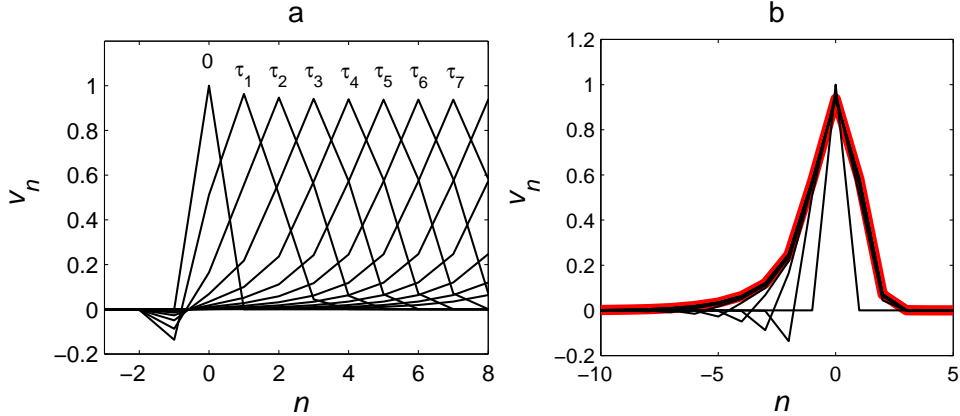}
\caption{Numerical simulations for the isothermal shell model with $\varepsilon = 0.5$. (a) Discrete dynamics induced 
by the transfer operator $\mathcal{T}_1$. Shown are the shell speeds $v_n$ 
at times $\tau_m$ corresponding to the wave positions $n_v(\tau_m) = m$ for $m = 0,1,2\ldots$ 
(b) After the shift $\mathcal{S}^m$ to the left, these solutions represent iterations of the Poincar\'e map 
$\mathcal{P}^m$ and converge to the fixed-point attractor (bold red curve) 
as $m \to \infty$.}
\label{fig2}
\end{figure}

In the isothermal shell model of previous section, solution (\ref{eq13}) corresponds to 
a fixed-point attractor of the Poincar\'e map,
see Fig.~\ref{fig2}. We will see that the case of nonzero shell temperatures yields variety of attractors depending on the parameter $\varepsilon$. Note that the approach proposed in this section is similar in spirit to the renormalization group method in partial differential equations, see, e.g.,~\cite{chen1995numerical,barenblatt1996scaling}.

\section{Bifurcation diagram}

\begin{figure}
\centering \includegraphics[width = 0.9\textwidth]{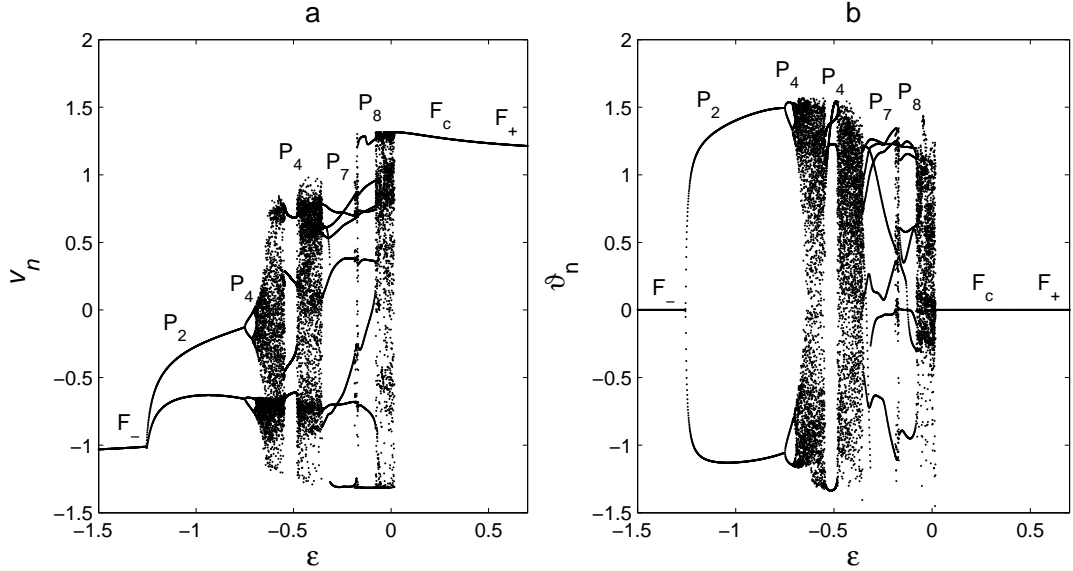}
\caption{Bifurcation diagram shows (a) the renormalized shell speeds $v_n$ and 
(b) the renormalized shell temperatures $\vartheta_n$ on the attractor of the Poincar\'e map for $n = n_v$. The fixed-point attractors $\mathrm{F}_+$, 
$\mathrm{F}_-$ and $\mathrm{F}_c$ 
are distinguished, see Section~\ref{sec_slefsim}. The largest periodic windows 
$\mathrm{P}_n$ are indicated,
which correspond to period-$n$ attractors of the Poincar\'e map.}
\label{fig3}
\end{figure}

We studied numerically the limiting behavior of a discrete dynamical system 
governed by the Poincar\'e map (\ref{eq30b}). 
For this purpose, we used 100 shells in system (\ref{eq7}), 
(\ref{eq8}). This gives a very good approximation, because the solution decays rapidly on both sides and occupies effectively about 10 shells, see Fig.~\ref{fig2}. 
Initial conditions with a few nonzero components centered at the shell $n_v = 70$ were chosen. 
In order to remove the transient behavior, the first $4000$ iterations were skipped. 
Figure~\ref{fig3} shows the bifurcation diagram represented by the speeds $v_n$ 
and temperatures $\vartheta_n$ of the shell $n_v = 70$.

Fixed-point (period-1) attractors exist for $\varepsilon < -1.255$. 
At $\varepsilon = -1.255$, the period-doubling bifurcation occurs leading to a period-2 attractor. 
The next period-doubling bifurcation corresponds to $\varepsilon = -0.749$, 
giving rise to the period-4 attractor. Further increase of $\varepsilon$ leads to 
the infinite period-doubling cascade, which follows the Feigenbaum 
scenario~\cite{feigenbaum1978quantitative} and ends with the chaotic behavior for $\varepsilon > -0.7$. 
For larger $\varepsilon$ a number of windows with periodic solutions are found,
which separate the regions with chaotic or quasi-periodic attractors. The largest periodic windows 
are indicated in Fig.~\ref{fig3}. 
At $\varepsilon = 0.015$ a subcritical bifurcation occurs 
(see Section \ref{sec_slefsim}) 
leading to a fixed-point attractor for $\varepsilon > 0.015$. 
Numerical results suggest that there is a unique attractor for each value of $\varepsilon$ (up to system symmetries), but multiple attractors may appear in general.

\begin{figure}
\centering \includegraphics[width = 0.96\textwidth]{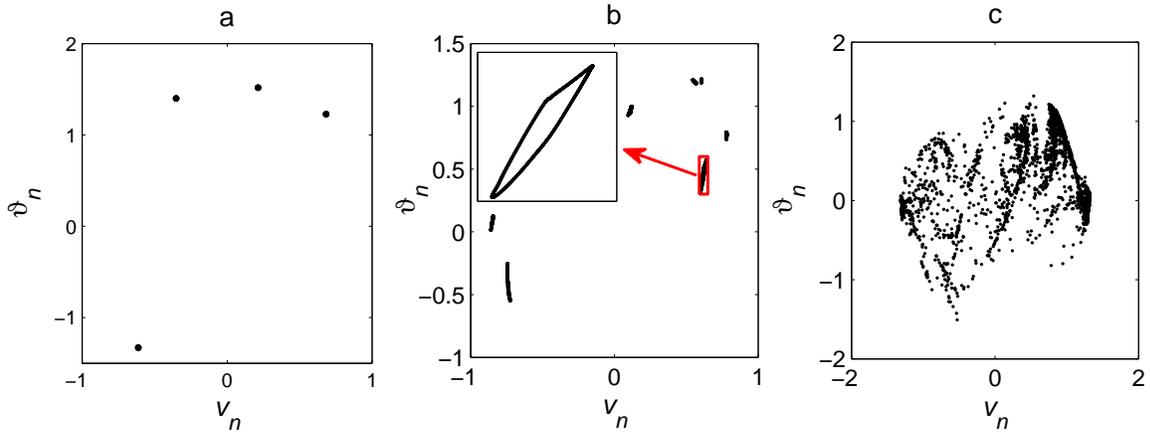}
\caption{Periodic, quasi-periodic and chaotic attractors on the plane $(v_n,\vartheta_n)$ 
for $n_v = 70$. These attractors correspond to $\varepsilon = -0.5$, $\varepsilon = -0.35$, and $\varepsilon = 0.01$, 
respectively.}
\label{fig4}
\end{figure}

Examples of periodic, quasi-periodic and chaotic attractors on the plane $(v_n,\vartheta_n)$ 
are shown in Fig.~\ref{fig4}. In the next sections we describe how each type of 
attractor determines the specific blowup structure.

\section{Self-similar blowup solutions}
\label{sec_slefsim}

\begin{figure}
\centering \includegraphics[width = 0.8\textwidth]{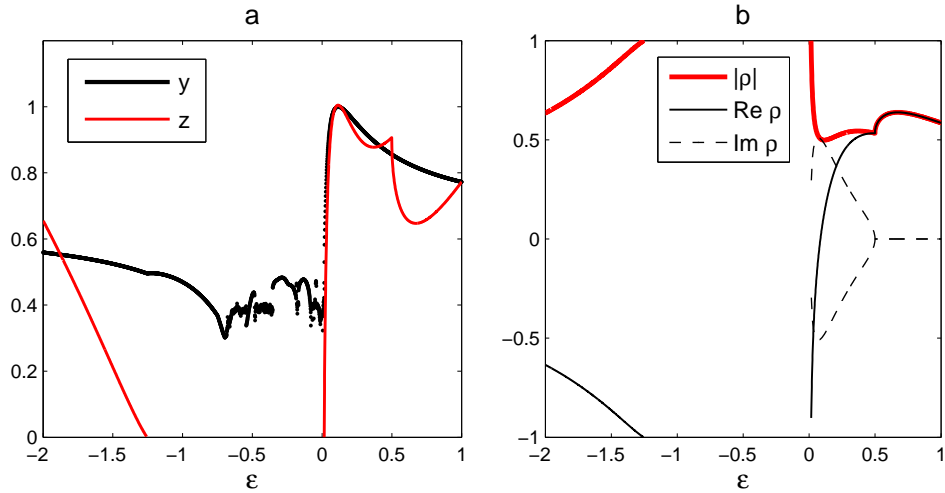}
\caption{(a) Scaling exponents $y$ (black) and $z$ (red) as functions of the shell model parameter $\varepsilon$. The exponent $z$ corresponds to fixed-point attractors. (b) The largest magnitude Floquet multiplier $\rho$ of the fixed-point attractor depending on $\varepsilon$.}
\label{fig5}
\end{figure}

First, let us consider a fixed-point (period-1) attractor $\mathbf{w}$ of the Poincar\'e map, i.e., 
\begin{equation}
\mathcal{P}\mathbf{w} = \mathbf{w}.
\label{eq41}
\end{equation}
In this case $\tau_1 = 1/a$, where $1/a$ is the time period determined by the Poincar\'e map. 
Using definition (\ref{eq30b}) for the shell speeds 
in (\ref{eq41}), we have 
\begin{equation}
v_{n+1}(\tau+1/a) = v_n(\tau).
\label{eq37}
\end{equation}
It is easy to see that a general sequence $v_n(\tau)$ satisfying condition (\ref{eq37}) has 
the form of a traveling wave (\ref{eq13}). 
The corresponding solution for the original variables $u_n(t)$ is 
described by Theorem~\ref{th2}.   
Figure~\ref{fig5}a shows the scaling exponent $y$ from (\ref{eq17}) 
as a function of $\varepsilon$ computed numerically 
(the values of $y$ corresponding to other types of attractors are discussed later). 
Since $y > 0$, the solution blows up at finite time $t_c$ given by (\ref{eq19}).

All shell temperatures vanish, $\vartheta_n = 0$, for the fixed-point attractor, 
as one can see from Fig.~\ref{fig3}b. In order to capture the limiting behavior of the original variables $\theta_n(t)$ 
near the blowup, we must consider small perturbations $\mathbf\delta\mathbf{w}(\tau)$ near 
the fixed-point $\mathbf{w}$. 
Following the classical perturbation theory (see, e.g.,~\cite{seyranian2003multiparameter}), 
we consider linearization of the Poincar\'e map near the fixed point, 
\begin{equation}
\mathcal{P}(\mathbf{w}+\delta\mathbf{w}) \approx \mathbf{w}
+d\mathcal{P}\,\delta\mathbf{w},
\label{eq39}
\end{equation}
where $d\mathcal{P}$ is the linear part (Jacobian matrix) of $\mathcal{P}$.
Let $\rho$ be 
the Floquet multiplier and $\widetilde{\mathbf{w}}$ be the corresponding eigenvector 
satisfying the eigenvalue problem
\begin{equation}
d\mathcal{P}\,\widetilde{\mathbf{w}} = \rho\widetilde{\mathbf{w}}.
\label{eq40}
\end{equation}
The multipliers are complex numbers in general and satisfy the condition $|\rho| < 1$ because $\mathbf{w}$ is an attractor. 
The asymptotic behavior of small perturbations $\delta\mathbf{w}(\tau)$ for large times is given by
\begin{equation}
\delta\mathbf{w}(\tau_m) \approx (d\mathcal{P})^m\delta\mathbf{w}(0)
\approx \rho^m \widetilde{\mathbf{w}},
\label{eq38}
\end{equation}
where $\tau_m = m/a$ and $\rho$ is the Floquet multiplier with maximum absolute value $|\rho|$. 
A linear combination must be taken in (\ref{eq38}) if there are several $\rho$ with equal maximum absolute values. 

Since  $\vartheta_n = 0$ for the fixed-point, the perturbed solution is 
$\vartheta_n = \delta\vartheta_n$ and we will drop $\delta$ below. 
The relations (\ref{eq30b}) are linear and, hence, are valid for the linearized 
operator $d\mathcal{P}$ as well. Therefore, similarly to (\ref{eq41}) and (\ref{eq37}), 
the eigenvalue equation (\ref{eq40}) for the shell temperatures yields
\begin{equation}
\vartheta_{n+1}(\tau+1/a) = \rho\,\vartheta_n(\tau).
\label{eq42}
\end{equation}
A general sequence $\vartheta_n(\tau)$ satisfying (\ref{eq42}) can be written in the form similar to (\ref{eq13}) as
\begin{equation}
\vartheta_n(\tau) = b\rho^n Q(n-a\tau),
\label{eq43}
\end{equation}
where $b$ is an arbitrary constant. 

The original shell variables $\theta_n(t)$ are expressed from
(\ref{eq43}) using the formulas (\ref{eq9}), (\ref{eq10}), where the quantities $\tau$ and $A$ correspond to the unperturbed (fixed-point) solution.  
The derivations analogous to those made in the proof of Theorem~\ref{th2} for shell speeds (\ref{eq13}) yield 
\begin{equation}
\theta_n(t) = b\rho^n k_n^{2y-1}\Theta(ak_n^y(t-t_c)).
\label{eq44}
\end{equation}
Assuming $a = 1$, the function $\Theta$ is determined by the expression similar to (\ref{eq22}) as
\begin{equation}
\Theta(t-t_c) = \exp\left(2\int_0^\tau A(\tau')d\tau'\right)Q(-\tau).
\label{eq45}
\end{equation}

The Floquet solutions and multipliers can be found numerically by standard methods, see, e.g., \cite{seyranian2003multiparameter}. The multiplier $\rho$ with 
maximum absolute value depending on parameter $\varepsilon$ is shown in Fig.~\ref{fig5}b. 
The multiplier is positive, negative or complex in three intervals denoted, 
respectively, by $\mathrm{F}_+ = [0.5,\,\infty)$, $\mathrm{F}_- = (-\infty,\,-1.255]$ and $\mathrm{F}_c = (0.017,\,0.5)$, see also Fig.~\ref{fig3}. In the tiny interval $0.015 < \varepsilon < 0.017$ (not shown in the figures), a real negative multiplier gets larger in magnitude than the complex ones. At $\varepsilon = 0.015$ a bifurcation occurs ($\rho = -1$). Since the fixed-point solution becomes unstable and no small-amplitude 
attractor appears for $\varepsilon < 0.015$, this bifurcation is subcritical, see, e.g.,~\cite{thompson1982instabilities}.

\begin{figure}
\centering \includegraphics[width = 0.98\textwidth]{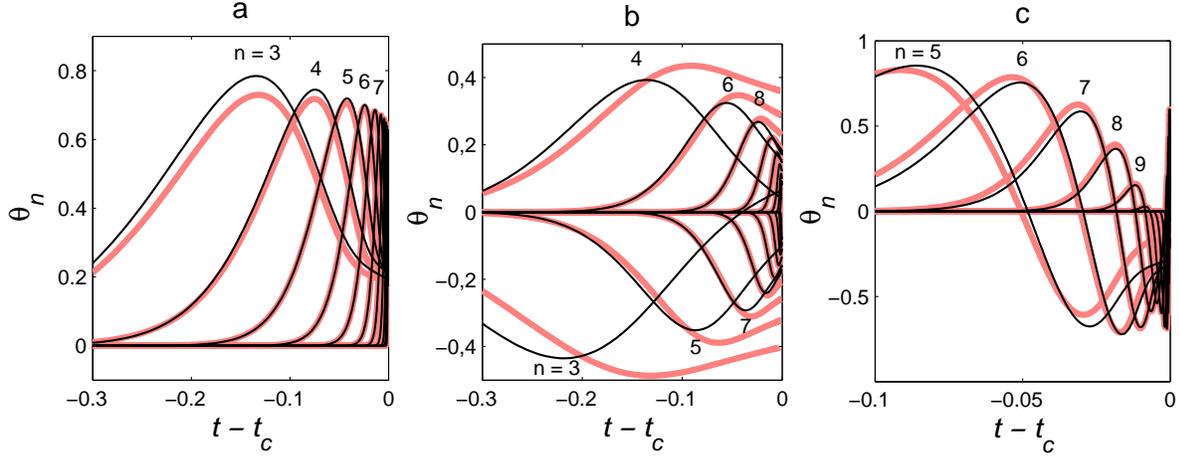}
\caption{Thin black curves show numerical solutions $\theta_n(t)$ for (a) $\varepsilon = 0.7$, (b) $\varepsilon = -1.5$, (c) $\varepsilon = 0.3$. These solutions are governed by the Floquet states with positive, negative and complex multipliers, respectively. Bold light-red curves represent the universal asymptotic form of the blowup given by (\ref{eq44a})--(\ref{eq44c}). In order to demonstrate convergence for large $n$, several initial shells are skipped.}
\label{fig6}
\end{figure}

We define
\begin{equation}
|\rho|^n = k_n^{-z},\quad z = -\log_h|\rho|, \quad \varphi = \arg \rho,
\label{eq46}
\end{equation}
where $z > 0$ due to the stability condition $|\rho| < 1$ mentioned above.
The scaling exponent $z$ depending on $\varepsilon$ is shown in Fig.~\ref{fig5}a.
For real positive $\rho$, expression (\ref{eq44}) takes the form
\begin{equation}
\rho > 0:\quad \theta_n(t) = bk_n^{2y-z-1}\Theta(ak_n^y(t-t_c)).
\label{eq44a}
\end{equation}
In the case of real negative $\rho$, we have
\begin{equation}
\rho < 0:\quad \theta_n(t) = b(-1)^nk_n^{2y-z-1}\Theta(ak_n^y(t-t_c)).
\label{eq44b}
\end{equation}
When $\rho$ is complex, the function $\Theta = \Theta_1+i\Theta_2$ 
and factor $b = b_1+ib_2$ are also complex, and the real solution 
is obtained by taking real part of (\ref{eq44}). This yields
\begin{equation}
\textrm{complex }\rho:\quad 
\theta_n(t) = k_n^{2y-z-1}
\mathrm{Re}\left[(b_1+ib_2)e^{in\varphi}(\Theta_1(\xi)+i\Theta_2(\xi))\right], 
\quad \xi = ak_n^y(t-t_c).
\label{eq44c}
\end{equation}
Figure~\ref{fig6} presents numerical results for $\varepsilon = 0.7$, $-1.5$ and $0.3$ corresponding to positive, negative and complex Floquet multipliers $\rho$. The numerical solutions $\theta_n(t)$ are shown (black curves) together with their asymptotic form (bold light-red curves) determined by expressions (\ref{eq44a})--(\ref{eq44c}). 

Since the fixed-point (\ref{eq41}) is an attractor, asymptotic solutions (\ref{eq13}) and (\ref{eq43}) are determined uniquely up to system symmetries (\ref{eq15})--(\ref{eqSym2}). For the original variables $u_n(t)$ and $\theta_n(t)$, these symmetries have the form (\ref{eq15n})--(\ref{eqSym2n}). One can check that these symmetries reduce to a change of the coefficients $a$ and $b$ in (\ref{eq21}) and (\ref{eq44a})--(\ref{eq44c})  only, leaving the scaling exponents $y$, $z$ and the functions $U$, $\Theta$ unaltered. We confirmed this fact numerically as well. 

We conclude that the fixed-point attractor of the Poincar\'e map $\mathcal{P}$ 
determines the asymptotic self-similar form of blowup. This form is described by expression (\ref{eq21}) for shell speeds and by (\ref{eq44a})--(\ref{eq44c}) for shell temperatures, where $a > 0$ and $b \in \mathbb{R}$ are arbitrary amplitudes. The scaling exponents $y$ and $z$, as well as the functions $U$ and $\Theta$ do not depend on initial conditions for a given shell model. 
Therefore, the asymptotic self-similar form of blowup is universal. 

\section{Blowup with periodic structure}

Let us consider a periodic attractor of the Poincar\'e map given by the vector $\mathbf{w}$ 
such that 
\begin{equation}
\mathcal{P}^p\mathbf{w} = \mathbf{w}, 
\label{eq60}
\end{equation}
where the period $p > 1$ is a positive integer. If $p/a$ is the time period of $\mathcal{P}^p$, then the times $\tau_n$ corresponding to $n$ iterations of the Poincar\'e map satisfy the conditions
\begin{equation}
\tau_n = \tau_j+pN/a, \quad n = j+pN, \quad 
j = 0,\ldots,p-1, \quad N = 0,1,2,\ldots
\label{eq61}
\end{equation}
Using (\ref{eq61}) and (\ref{eq30b}), periodicity condition (\ref{eq60}) can be written as
\begin{equation}
v_{n+p}(\tau+p/a) = v_n(\tau), \quad 
\vartheta_{n+p}(\tau+p/a) = \vartheta_n(\tau). 
\label{eq62}
\end{equation}
A general form of the functions satisfying (\ref{eq62}) is
\begin{equation}
v_n(\tau) = aV_j(n-a\tau), \quad 
\vartheta_n(\tau) = a^2Q_j(n-a\tau), \quad j = 0,\ldots,p-1,
\label{eq63}
\end{equation}
where $n = j+pN$ and the factors $a$ and $a^2$ are used to reflect the symmetry (\ref{eq15}). 
For each $j$ solution (\ref{eq63}) represents a wave moving with speed $a$ along the logarithmic axis $n = \log_h k_n$. 

\begin{theorem}
Taking $a = 1$ in (\ref{eq63}) we define the real quantity
\begin{equation}
y = \frac{1}{p\log h}\int_0^{p}A(\tau) d\tau
\label{eq90}
\end{equation}
and the functions
\begin{eqnarray}
U_j(t-t_c) & = & k_j^{-y}\exp\left(\int_0^\tau A(\tau')d\tau'\right)V_j(j-\tau),
\label{eq91}
\\
\Theta_j(t-t_c) & = & k_j^{-2y}\exp\left(2\int_0^\tau A(\tau')d\tau'\right)Q_j(j-\tau),
\label{eq91b}
\end{eqnarray}
where $\tau$ is related to $t$ by (\ref{eq9}) and $A$ is given by (\ref{eq12}). If $y > 0$, then the solution $u_n(t)$, $\theta_n(t)$ corresponding to (\ref{eq63}) has the form (for arbitrary $a > 0$) 
\begin{equation}
u_n(t) = ak_n^{y-1}U_j(ak_{n-j}^y(t-t_c)), \quad
\theta_n(t) = a^2k_n^{2y-1}\Theta_j(ak_{n-j}^y(t-t_c)),
\label{eq92}
\end{equation}
and describes the blowup at $t_c$ given by (\ref{eq19}).
\label{th3}
\end{theorem}

\noindent\textit{Proof:}
The coefficient $a$ in (\ref{eq63}) and (\ref{eq92}) reflects the time-scaling symmetries (\ref{eq15}) and (\ref{eq15n}). Hence, we can take $a = 1$ in the rest of the proof. 
We will derive expressions (\ref{eq91}) and (\ref{eq92}) for the shell speeds; the shell temperatures in (\ref{eq91b}) and (\ref{eq92}) are considered similarly.

The function $A(\tau)$ given by (\ref{eq12}) and (\ref{eq63}) is periodic  with period $p$. Due to (\ref{eq90}) we have
\begin{equation}
\exp\left(\int_\tau^{\tau+pN}A(\tau') d\tau'\right) 
= k_{pN}^{y}
\label{eq93}
\end{equation}
for any $\tau$, where $k_{pN} = h^{pN}$.
Let us consider the time $t'$ corresponding to $\tau+pN$. 
Using (\ref{eq9}) and (\ref{eq19}), the derivation analogous to (\ref{eq34}) yields
\begin{equation}
t_c-t' = k_{pN}^{-y}(t_c-t).
\label{eq94}
\end{equation}
From (\ref{eq63}) we find $v_n(\tau+pN) = V_j(n-\tau-pN) = V_j(j-\tau)$ for $n = j+pN$.
Then we express the shell speeds $u_n$ from (\ref{eq10}) and use (\ref{eq93}), (\ref{eq91}) as
\begin{equation}
\begin{array}{rcl}
u_n(t') 
& = & \displaystyle
k_n^{-1}\exp\left(\int_0^{\tau+pN} A(\tau') d\tau'\right)v_n(\tau+pN) 
\\[15pt]
& = & \displaystyle
k_n^{-1}k_{pN}^{y}\exp\left(\int_0^{\tau} A(\tau') d\tau'\right)V_j(j-\tau) = k_n^{y-1}U_j(t-t_c).
\end{array}
\label{eq95}
\end{equation}
Substituting $t-t_c = k_{n-j}^y(t'-t_c)$ expressed from (\ref{eq94}) into (\ref{eq95}) and dropping the prime, we obtain the first formula in (\ref{eq92}) with $a = 1$. 

It follows from periodicity of $A(\tau)$ and (\ref{eq90}) that the inequality (\ref{eq18}) holds for some constant $C$, which implies convergence of the integral in (\ref{eq19}) and the blowup of solution (\ref{eq92}) at $t_c$ for $y > 0$, just as in the proof of Theorem~\ref{th2}.
$\square$

\begin{figure}
\centering \includegraphics[width = 0.8\textwidth]{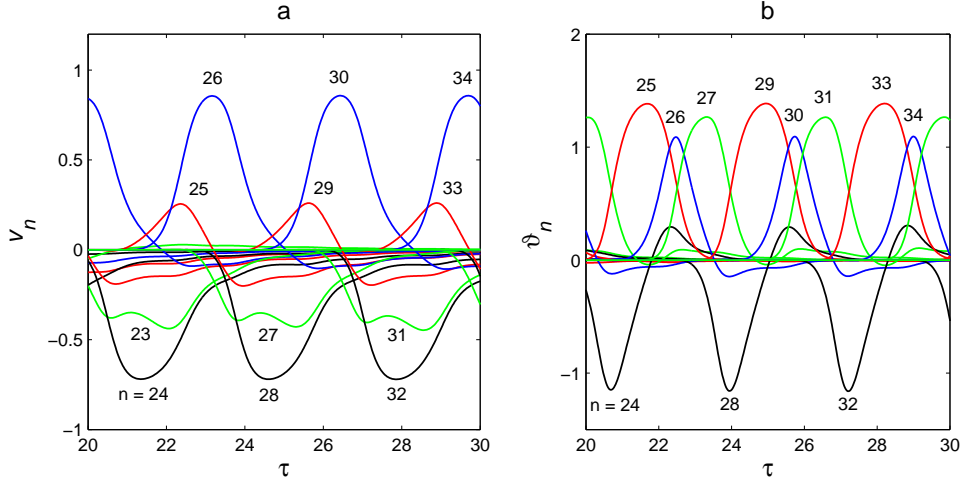}
\caption{Asymptotic form of renormalized shell speeds $v_n(\tau)$ and temperatures $\vartheta_n(\tau)$ for the blowup with periodic structure (\ref{eq63}) in the shell model with $\varepsilon = -0.5$. This solution corresponds to the $\mathrm{P}_4$ window in Fig.~\ref{fig3}.}
\label{fig7}
\end{figure}

Since $\vartheta_n \ne 0$ in the periodic attractor, the speed and temperature variables are strongly coupled in the blowup, in contrast to the case of the fixed-point attractor in Section~\ref{sec_slefsim}. As a consequence, there is a single scaling exponent $y$ in self-similar expressions of Theorem~\ref{th3}. Periodic solution (\ref{eq60}) is an attractor and, hence, expressions (\ref{eq63}) represent the asymptotic form of solution in renormalized variables for large $\tau$, Fig.~\ref{fig7}.
The functions $V_j$ and $Q_j$ in (\ref{eq63}) are determined uniquely up to system symmetries. The symmetry (\ref{eq15}) is taken into account in (\ref{eq63}), and the symmetries  (\ref{eqSym}), (\ref{eqSym2}) reduce to the shift of time $\tau$ and the cyclic permutations of the indexes $j \mapsto j+1\ (\mathrm{mod}\ p)$. 

\begin{figure}
\centering \includegraphics[width = 0.8\textwidth]{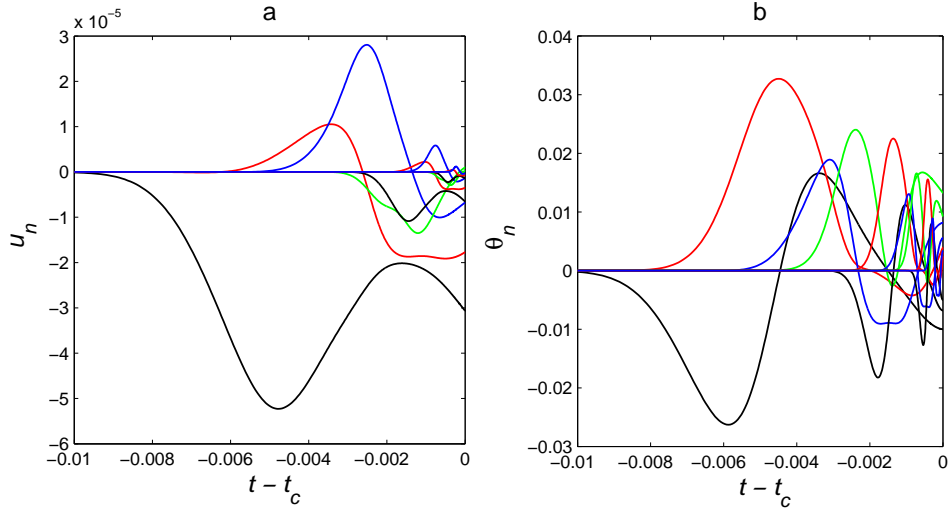}
\caption{Shell velocities $u_n(t)$ and temperatures $\theta_n(t)$ for the shell model with $\varepsilon = -0.5$. The shells $n = 24,\ldots,34$ are shown, which correspond to the solution displayed in Fig.~\ref{fig7}. The solution blows up as $t \to t_c^-$ and has the period-4 structure visible in the figure.}
\label{fig8}
\end{figure}

We conclude that the blowup with periodic structure has asymptotic form (\ref{eq92}) for large $n$ and $t \to t_c^-$, where the scaling exponent $y$ and the functions $U_j$, $\Theta_j$ are universal, i.e., independent of initial conditions (up to cyclic permutations of indexes $j = 0,\ldots,p-1$).
The value of $y$ depending on the parameter $\varepsilon$ is shown in Fig.~\ref{fig5}a. Figure~\ref{fig8} presents a numerical example of the  blowup.

\section{Blowup with quasi-periodic and chaotic structure}

\begin{figure}
\centering \includegraphics[width = 0.8\textwidth]{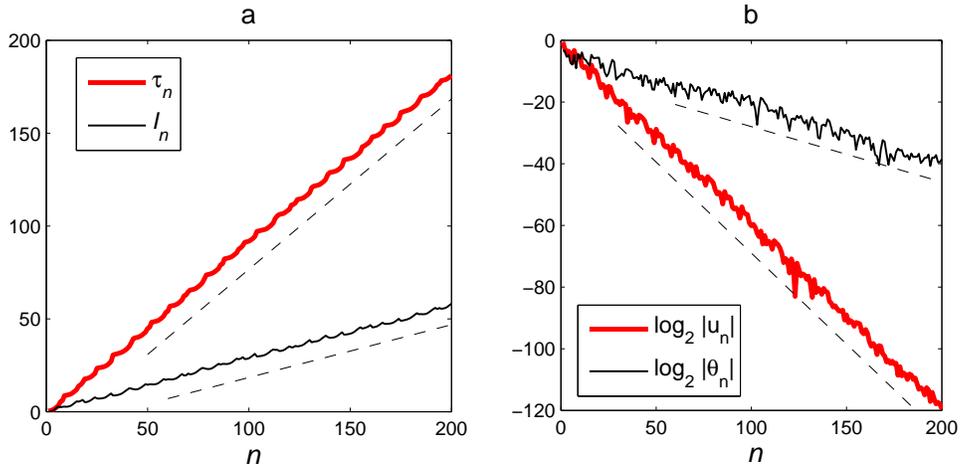}
\caption{(a) The times $\tau_n$ and integrals $I_n = \int_0^{\tau_n}A(\tau)d\tau$ corresponding to $n$ iterations of the Poincar\'e map $\mathcal{P}$ for $\varepsilon = 0.01$ (chaotic attractor). The dashed lines indicate the slopes $1/a$ and $\langle A\rangle/a = y\log h$. (b) Power law scaling of shell speeds $u_n(t)$ and temperatures $\theta_n(t)$ corresponding to the times when $n_v = n$. The dashed lines indicate the slopes $y-1$ and $2y-1$ given by (\ref{eq96}) and (\ref{eq98}).}
\label{fig9}
\end{figure}

According to numerical simulations in Figs.~\ref{fig3} and \ref{fig4} quasi-periodic and chaotic attractors exist in the interval $-0.7 < \varepsilon < 0.015$, outside the periodic windows. In particular, the value $\varepsilon = 0.01$ used in~\cite{brandenburg1992energy,Ching2010} corresponds to the chaotic attractor. Figure~\ref{fig9}a shows the times $\tau_n$ and the integrals $\int_0^{\tau_n} A(\tau)d\tau$ corresponding to $n$ iterations of the Poincar\'e map for $\varepsilon = 0.01$. These quantities grow linearly with $n$ up to small chaotic oscillations. 

We introduce the quantities
\begin{equation}
\frac{1}{a} = \lim_{n \to \infty}\frac{\tau_n}{n} > 0,\quad
\langle A\rangle = \lim_{n \to \infty}\frac{1}{\tau_n}\int_0^{\tau_n} A(\tau) d\tau,\quad
y = \frac{\langle A\rangle}{a\log h},
\label{eq25a}
\end{equation}
where $1/a$ represents the mean time step $\langle \tau_n-\tau_{n-1} \rangle$ of the Poincar\'e map and $\langle A\rangle$ is the mean value of the function $A(\tau)$ on the attractor. Note that an arbitrary value $a > 0$ can be obtained after the time-scaling symmetry transformation (\ref{eq15}), but this transformation does not alter the value of $y$. Also, for fixed-point and periodic attractors, the last expression in (\ref{eq25a}) reduces to (\ref{eq17}) and (\ref{eq90}). 

Since $A(\tau) = \langle A\rangle+\delta A(\tau)$, where $\delta A$ oscillates near zero mean value, the inequality (\ref{eq18}) written for $a = 1$ holds, see Fig.~\ref{fig9}a. As a result, the integral in (\ref{eq19}) converges for $y > 0$ providing the finite blowup time $t_c$.

\begin{figure}
\centering \includegraphics[width = 0.8\textwidth]{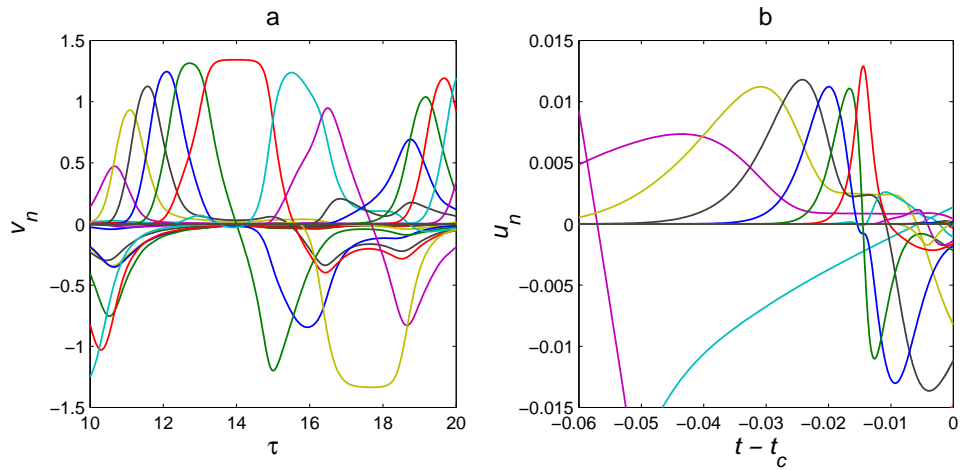}
\caption{(a) Dynamics of renormalizes variables $v_n(\tau)$ represents a chaotic wave moving with average speed $a$ toward large shells $n$. (b) The corresponding solution for original shell variables $u_n(t)$ blows up in finite time $t_c$. The blowup is hard to recognize visually. However, it satisfies the scaling laws (\ref{eq96})--(\ref{eq99}) with the universal exponent $y$, see Fig.~\ref{fig9}b.}
\label{fig10}
\end{figure}

Figure~\ref{fig10}a shows the speeds $v_n(\tau)$, whose dynamics can be viewed as a chaotic wave moving toward large shell numbers $n$. 
Since one iteration of the Poincar\'e map corresponds to the increase of wave position by one, the mean wave speed equals $a$. The wave amplitude in renormalized variables $v_n$ and $\vartheta_n$ does not change, see (\ref{eq12b}). Using (\ref{eq25a}) we estimate
\begin{equation}
\exp\left(\int_0^{\tau_n} A(\tau') d\tau'\right) 
= e^{\langle A\rangle\tau_n}\exp\left(\int_0^{\tau_n} \delta A(\tau') d\tau'\right) 
\sim e^{\langle A\rangle\tau_n}
\sim e^{\langle A\rangle n/a} \sim k_n^y.
\label{eq97}
\end{equation}
Therefore, when the wave reaches the shell $n$, the shell speeds and temperatures can be estimated by order of magnitude using (\ref{eq10}) as 
\begin{eqnarray}
u_n & = & k_n^{-1}\exp\left(\int_0^{\tau_n} A(\tau') d\tau'\right)v_n \sim k_n^{y-1},
\label{eq96}
\\
\theta_n & = & k_n^{-1}\exp\left(2\int_0^{\tau_n} A(\tau') d\tau'\right)\vartheta_n
\sim k_n^{2y-1}.
\label{eq98}
\end{eqnarray}
The corresponding time interval to the blowup is found using (\ref{eq9}), (\ref{eq19}), (\ref{eq25a}) and (\ref{eq97}) as
\begin{equation}
\begin{array}{rcl}
t_c-t 
& = & \displaystyle
\int_{\tau_n}^\infty \exp\left(-\int_0^{\tau'} A(\tau'') d\tau''
\right)d\tau' 
= \int_0^\infty \exp\left(-\int_0^{\xi'+\tau_n} A(\tau'') d\tau''
\right)d\xi' 
\\[12pt] & = & \displaystyle
\exp\left(-\int_0^{\tau_n} A(\tau'') d\tau''\right)
\int_0^\infty \exp\left(-\int_{0}^{\xi'} A(\xi''+\tau_n) d\xi''
\right)d\xi' 
\sim k_n^{-y},
\end{array}
\label{eq99}
\end{equation}
where we changed the integration variables as $\tau' = \xi'+\tau_n$ and $\tau'' = \xi''+\tau_n$; note that the last integral factor represents an oscillating quantity with finite mean value.

Scaling laws (\ref{eq96})--(\ref{eq99}) are the same as for the blowup with periodic structure, see (\ref{eq92}). They are confirmed numerically in Fig.~\ref{fig9}b, which presents the logarithmic plots of the amplitudes $|u_n(t)|$ and $|\theta_n(t)|$  at the times $t$ when the wave position $n_v$ is equal to $n$. 
The scaling exponent $y$ computed numerically is shown in Fig.~\ref{fig5}a. Since $y > 0$, the estimate (\ref{eq96}) shows that $k_nu_n$ becomes infinitely large as $t \to t_c^-$ confirming the blowup at $t_c$ for all values of the shell model parameter $\varepsilon$. Figure~\ref{fig10}b gives an example of the blowup solution $u_n(t)$ for $\varepsilon = 0.01$. 

Therefore, in the case of quasi-periodic and chaotic attractors, solutions $u_n(t)$, $\theta_n(t)$ blow up in finite time. The asymptotic structure of these solutions near the blowup is universal, i.e., they scale as (\ref{eq96})--(\ref{eq99}) with the single universal exponent $y$ and the dynamics of renormalized variables $v_n(t)$, $\vartheta_n(t)$ is described asymptotically by the attractor of the Poincar\'e map. 

\section{Conclusion}

We considered the inviscid shell model of convective turbulence suggested in~\cite{brandenburg1992energy} and studied the blowup problem, i.e., formation of a finite-time singularity for smooth solutions of finite norm. The blowup condition similar to the Beale-Kato-Majda theorem is derived. The constructive method for analysis of blowup is developed, which exploits the Poincar\'e map introduced for a renormalized system. This method identifies the blowup and its internal structure by studying attractors of the Poincar\'e map.

It is shown that solutions of the shell model blow up in finite time for all values of the parameter, which controls the process of energy transfer. Depending on the value of this parameter, fixed-point, periodic, quasi-periodic and chaotic attractors of the Poincar\'e map were detected, and the corresponding structure of the blowup was described. In each case the asymptotic form of blowup turns out to be universal (independent of initial conditions) and described by the specific scaling law.

A similar renormalization procedure can be used for continuous hydrodynamic models. For the inviscid Burgers equation, an asymptotic solution for renormalized Fourier variables represents a traveling wave~\cite{mailybaev2012} and describes the universal self-similar form of blowup~\cite{eggers2009role}. This construction is analogous to the traveling wave corresponding to a fixed-point attractor, which determines self-similar blowup structures in this paper. Our analysis demonstrates that the blowup structure can be quite sophisticated and hard to recognize, e.g., the high-period or chaotic blowup structure. Thus, using special renormalization techniques is crucial. This knowledge may be useful for studying the open problems of blowup in incompressible Euler equations and fully inviscid Boussinesq equations. 

The obtained results can also be useful for explaining spectra of developed turbulence 
in the inertial range.  It was shown in~\cite{mailybaev2012computation} that the blowup phenomenon 
is the source for anomaly of turbulent spectra (deviations from scaling exponents given by the phenomenological Kolmogorov theory~\cite{frisch1995turbulence}) in the Sabra shell model. In this case self-similar blowup structures develop in the inertial range and dissipate in the viscous range, 
leading to coherent events called the instantons. For the shell model considered in the present paper, analysis of instanton statistics in turbulent dynamics is a difficult problem due to nontrivial, e.g., chaotic blowup structure. In this case, identification of the instantons must rely on the knowledge of the blowup structure.

Since scaling properties of the blowup is the important issue in the above applications, we summarize that there are two types of power scaling laws describing the blowup in the shell model of convective turbulence. The first type corresponds to the interval $-1.255 < \varepsilon < 0.015$ (periodic, quasi-periodic and chaotic attractors) and described by the power laws
\begin{equation}
u_n \sim k_n^{y-1},\quad
\theta_n \sim k_n^{2y-1},\quad
t \sim k_n^{-y}
\label{eq100}
\end{equation}
with a single exponent $y$. Velocity and temperature fields are essentially coupled in this case. For $y = 2/5$ expressions (\ref{eq100}) reduce to the Bolgiano-Obukhov scaling law based on dimensional analysis. Since $y \ne 2/5$ for most values of $\varepsilon$, see Fig.~\ref{fig5}a, the blowup scaling is anomalous. 

The second type corresponds to the intervals $\varepsilon < -1.255$ and $\varepsilon > 0.015$ (fixed-point attractors). It is described by the power laws
\begin{equation}
u_n \sim k_n^{y-1},\quad
\theta_n \sim k_n^{2y-z-1},\quad
t \sim k_n^{-y}
\label{eq101}
\end{equation}
with two independent exponents $y$ and $z$. The velocity and temperature fields uncouple in this case: the scaling exponent $y$ is determined by the velocity field only, and the exponent $z$ describes small temperature perturbations. 

\section*{Acknowledgments} 
The author is grateful to Edriss S. Titi for useful discussions. 
This work was supported by CNPq under grants 477907/2011-3 and 305519/2012-3.

\bibliographystyle{plain}
\bibliography{refs}

\end{document}